\documentclass[aps,prl,preprint,showpacs,preprintnumbers,floatfix]{revtex4}

\usepackage{graphicx}

\begin{document}

\title{Precise Barriers and Shell Effects: a New Inroad to Fission Saddle Point Spectroscopy}

\author{L. Phair, L.~G. Moretto, K.~X. Jing, L. Beaulieu,
        D. Breus, \\ J.~B. Elliott, T.~S. Fan, Th. Rubehn, and
        G.~J. Wozniak} 
\affiliation{Nuclear Science Division, Lawrence
        Berkeley National Laboratory, University of
        California, Berkeley, California 94720}

\date{\today}

\begin{abstract}
   Fission excitation functions have been measured for a
   chain of neighboring compound nuclei, from $^{207}$Po to
   $^{212}$Po. We present a new analysis which provides a
   determination of the fission barriers and ground state shell
   effects with nearly spectroscopic accuracy.  The improved accuracy achieved in this analysis may lead
   to a future detailed exploration of the saddle mass surface and its spectroscopy. The sensitivity of the fission probabilities on shell effects extends to excitation energies of 150 MeV and negates recent claims for the disappearance of shell corrections due to collective effects.
\end{abstract}


\preprint{LBNL-51709}

\maketitle

The study of nuclei under extreme conditions (spin, isospin,
temperature, and deformation) is a major theme of nuclear
physics. Fission is a fertile testing ground
of nuclei under extreme deformation for several reasons.

A fissioning nucleus allows exploration of the {\em most} extreme nuclear
deformation associated with a stationary point, beyond that of
super- or even hyper-deformation.  The saddle configuration
is a bottleneck in phase space, a ``stationary'' point at
which the probability to fission is determined. It is able to sustain
its own spectroscopy in the $N-1$ modes orthogonal to the fission
mode. This spectroscopy begins with the saddle mass ($M_s$) which is the
ground state mass ($M_{gs}$) plus the experimental fission barrier
($B_f$) \cite{Swiatecki99}. 
Initial attempts at saddle point spectroscopy were made earlier in the low barrier actinide regions \cite{some_review}, but could not be extended to the  higher barriers of lighter elements.

Historically, experimental fission barriers in lighter nuclei have been disproportionately useful
in fixing the adjustable parameters in theories of nuclear masses and
deformabilities (such as the liquid drop model).

Shell effects in the ground state and at the saddle, pairing, congruence energy \cite{congruence}, single particle level densities are examples of quantities that should be immediately accessible when studying saddle properties. 

Yet as important a testing ground as fission would seem to be,
fission barriers have been measured only anecdotally and
with moderate accuracy. The lack of precise and systematic data
measured over a wide range of excitation energy has left the
expectations mentioned above largely unfulfilled.

In this letter we provide new precision data, systematically measured
for an isotopic chain of Po compound nuclei, covering a large
excitation energy range. 
We describe a new analysis which results in fission barriers and ground state
shell corrections with nearly spectroscopic accuracy.
Using this method we have measured fission barriers and saddle masses with a precision 10 times greater than anything achieved before \cite{Moretto95b}, opening the possibility of determining subtle and important features of the saddle point. 
 In the process, we have also measured accurately ground state shell corrections by measuring fission probabilities. Since the shell corrections we extract are accurate  (we determine this from independent data), we have confidence that the fission barriers have a similar degree of accuracy. If we measure enough of these fissioning systems we will
be able to determine the fission saddle mass surface as a function of $Z, A$, fissility, etc. Indeed,   the spectroscopy of the fission saddle point will soon be open to us.

\begin{figure}
\centerline{\includegraphics[,width=8.0cm,angle=0]{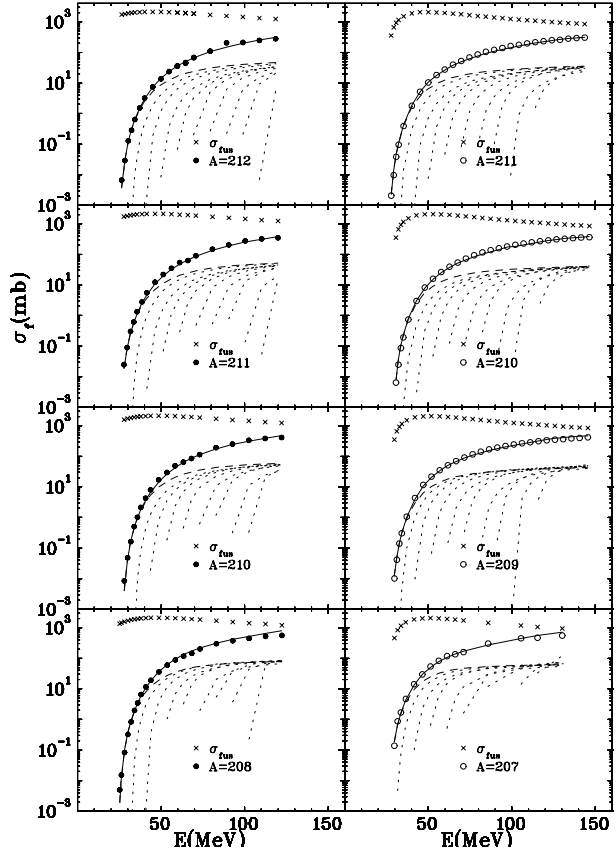}}
\caption{The fission cross section (solid and open symbols) are plotted as a function
of excitation energy for the indicated nuclei. The errors are smaller than the symbols. The dashed
curve represents the first chance fission cross section. The dotted
curves represent the second and higher chance fission cross
sections. The solid curve is their sum, the total fission cross
section. The left column contains $\alpha$-induced reactions. The right contains $^{3}$He-induced reactions. The fusion cross sections (cross symbols) are described in the text.}
\label{fig:fits}
\end{figure}

The fission data were taken at the 88-Inch Cyclotron of the Lawrence
Berkeley National Laboratory.  We measured with high precision the
fission excitation functions of the neighboring compound Po
nuclei $^{207-212}$Po produced in $^{3}$He- and
$^{4}$He-induced reactions on isotopically enriched Pb targets (see
Fig.~\ref{fig:fits}). 

We chose these particular reactions for several reasons. First, the
shell corrections and fission barriers in the Pb region are large.
Second, the light ion induced reactions have
only modest amounts of angular momentum ($<25\hbar$). The relevant
rotational energies are small, $\approx$ 2 MeV for a spherical
shape and $\approx 0.8$ MeV for the saddle shape of a Po nucleus with
an angular momentum of 20$\hbar$. And third, there are four stable
isotopes of Pb from which one can make clean targets.

Fission events were identified in two large area parallel plate
avalanche counters. The experimental details are described in
ref.~\cite{Rubehn96}. The solid and open symbols in Fig.~\ref{fig:fits} represent the fission cross
section data for neighboring compound nuclei. 
 The fission cross sections cover seven orders of magnitude for these reactions. In two cases ($A=211,210$), we have
overlap points where the same compound nucleus was formed via two
different entrance channels.
    
To determine the fission probability, we use
standard transition state theory 
\cite{Rubehn96} to calculate the fission width
\begin{equation}
\Gamma _f= \frac{1}{2\pi\rho (E)}\int\rho _f(E-B_f-\epsilon )d\epsilon
\label{eq:Gamma_f}
\end{equation}
where $\rho _f$ is the level density at the saddle, $\epsilon$ is the kinetic energy associated with the fission channel, and $\rho$ is the
level density of the compound nucleus.

We neglect charged particle emission, since fission following proton or $\alpha$ particle emission is known to be small for these reactions \cite{Jing:tau_D}.
The width for neutron emission (the only other exit channel assumed in
our analysis)
is 
\begin{equation}
\Gamma_n=\frac{2mR^2g\prime}{\hbar^2}\frac{1}{2\pi\rho
(E)}\int\epsilon\rho _d(E-B_n-\epsilon )d\epsilon .
\label{eq:Gamma_n}
\end{equation}
where $m$ denotes the neutron mass, $R$ is the radius and $\rho _d$ is
the level density of the daughter nucleus after neutron emission,
$g\prime$ is the spin factor (=2), $B_n$ is the neutron
binding energy, and $\epsilon$ is the kinetic energy of the neutron.



Using for simplicity the Fermi gas level density and taking into
 account the angular momentum a fissioning nucleus may have,
 Eqs. (\ref{eq:Gamma_f}) and (\ref{eq:Gamma_n}) can be evaluated and
 their ratio taken \cite{Jing99Thesis} so that $\Gamma_f/\Gamma_n$ is
\begin{equation}
\frac{\Gamma_f}{\Gamma_n}=\frac{T_f-\frac{1}{2a_f}}{K\left(
T_d^2-\frac{3}{2a_d}T_d+\frac{3}{4a_d^2}\right)}\frac{\rho_f(E-B_f-E_r^s)}{\rho_d
(E-B_n-E_r^{gs})}
\label{eq:exact}
\end{equation}
where $a_f$ and $T_f$ denote the level density parameter and
temperature at the saddle, $a_d$ and $T_d$ denote the same quantities for the residual daughter after neutron
emission, $E_r^s$ and $E_r^{gs}$ denote the rotational energy of
the system at the saddle point and the energy of the rotating ground
state and $K=2mR^2g\prime / \hbar ^2$.
The ground state and saddle
moments of inertia were taken from Sierk \cite{Sierkref}.

Using for simplicity the Fermi gas level density, the ratio of the level densities in Eq.~(\ref{eq:exact})
becomes
\begin{equation}
\frac{\rho_f(E-B_f-E_r^s)}{\rho_d(E-B_n-E_r^{gs})}\propto
e^{2\sqrt{a_f(E-B_f-E_r^s)}-2\sqrt{a_d(E-B_n-E_r^{gs})}}.
\end{equation}

For nuclei with strong shell effects,
the approximation $\rho_d(E-B_n-E_r^{gs})\propto\exp
(2\sqrt{a_d(E-B_n-E_r^{gs})})$ becomes a poor one. The shell effects
of a nucleus affect its level density in a rather complicated way at
low energies. But at high enough excitation energies, we can use the
asymptotic form $\rho (E)\propto\exp (2\sqrt{a(E+\Delta_{\rm
shell})})$ \cite{Hui72}. This approximation is particularly useful
for $\Gamma _n$ where the excitation energy is at least 5 MeV
above the fission barrier (i.e., $\approx$ 25-30 MeV). On the
other hand, at the saddle no shell effects are expected much larger
than 1 MeV.

For the daughter nucleus, $\rho_d$ takes the asymptotic form:
\begin{equation}
\rho _d(E-B_n-E_r^{gs})\propto
\exp\left(
2\sqrt{a_d(E-B_n-E_r^{gs}+\Delta_{\rm shell}^{n-1})} \right)
\end{equation}
where $\Delta _{\rm shell}^{n-1}$ is the ground state shell effect of
the daughter nucleus after neutron emission.

For the saddle level density ($\rho _f$), the problems
should be less serious. First, large deformations at
the saddle point imply small shell effects there. And second, the saddle point
should be topologically located between regions of positive and negative shell effects, thus substantially limiting the saddle point shell corrections \cite{Myers96}.
   
Pairing affects the level density in a manner similar to
 the shell effects. The level density is evaluated at an energy
shifted by the condensation energy $\Delta E_c$. The condensation
energies are calculated separately for protons and neutrons. For an
even-even nucleus,
$\Delta E_c=\frac{1}{2}g_n\Delta _{n}^2 + \frac{1}{2}g_p\Delta _{p}^2$, 
where $g_n=(3/\pi^2)a_n$, $g_p=(3/\pi^2)a_p$, and
$a_d=a_n+a_p=N/8.5~{\rm MeV}^{-1}+Z/8.5~{\rm MeV}^{-1}=A/8.5~{\rm
MeV}^{-1}$.  
In general,
\begin{equation}
\Delta E_c=\frac{1}{2}g_n\Delta _{n}^2+\frac{1}{2}g_p\Delta _{p}^2 -
{\rm mod}(N,2)\Delta_{n} - {\rm mod}(Z,2)\Delta _{p}.
\end{equation}
The ground state gap parameters for protons ($\Delta _{p}$) and for
neutrons ($\Delta _{n}$) were chosen to be
\begin{equation}
\Delta _{p}=\Delta _{n}=\frac{12 {\rm MeV}}{\sqrt{A}}.
\label{eq:sqrt12}
\end{equation}

At the saddle, the gap parameter for the neutrons($\Delta _{n}^f$) was
taken to be
$\Delta _{n}^f=S\exp\left({-{1}/{g_{n}^{f}G}}\right)$
where $S$ and $G$ were chosen to reproduce the ground state values
(Eq.~(\ref{eq:sqrt12})) and $g_{n}^{f}=(3/\pi^2)(N/A)a_f$. A similar
expression for $\Delta _{p}^f$ can be calculated for protons using
$g_{p}^{f}=(3/\pi^2)(Z/A)a_f$. Consequently the condensation energy at
the saddle we express as
\begin{equation}
\Delta E_c^s=\frac{1}{2}g_{n}^{f}(\Delta
_{n}^f)^2+\frac{1}{2}g_{p}^{f}(\Delta _{p}^f)^2 \\ -{\rm
mod}(N,2)\Delta _{n}^f-{\rm mod}(Z,2)\Delta _{p}^f
\end{equation} 
The resulting expression for $\Gamma _f/\Gamma _n$ is
\begin{equation}
\frac{\Gamma _f}{\Gamma _n}\propto e^{2\sqrt{a_f\left(
E-B_f-E_r^s-\Delta E_c^f\right)}-2\sqrt{a_d\left(E-B_n-E_r^{gs}+\Delta
_{\rm shell}^{n-1}-\Delta E_c\right)}}.
\label{eq:gammaf_gamman}
\end{equation}
We further assume that the fission barrier has two
parts: 
$B_f=B_{\rm macro} - \Delta _{\rm shell}$.
For the macroscopic part ($B_{\rm macro}$) we take a scaled
value of the Thomas-Fermi predictions \cite{Swiatecki99}. The
microscopic part is the ground state shell correction.


The expression for $\Gamma _f/\Gamma _n$
(Eq.~(\ref{eq:gammaf_gamman})) has four free parameters: $B_{\rm
macro}$, $\Delta _{\rm shell}$ of the compound system, $\Delta _{\rm
shell}^{n-1}$ of the 1 neutron out daughter nucleus, and the ratio $a_f/a_d$.
To use this description of $\Gamma_f/\Gamma_n$, we write the
total fission cross section as
\begin{equation}
\sigma _f=\sum _{i=0}\sigma _f^{(i)}=\sum _{l=0}^{l=l_{\rm max}}\sum
_{i=0}\sigma _l P_f^{(i)}(E,l)
\label{eq:multichance}
\end{equation} 
where $\sigma _f^{(i)}$ is the fission cross section after $i$
neutrons have been emitted, $\sigma _l$ is the angular momentum
distribution of the fusion cross section ($(2l+1)\pi\lambdabar ^2$),
$l_{\rm max}$ comes from the fusion cross sections (crosses in
Fig.~\ref{fig:fits}) and $P_f^{(i)}(E,l)$ is the fission probability
after the emission of $i$ neutrons from a compound nucleus of initial
angular momentum $l$ and initial energy $E$. The fission probability
at each ``step'' $i$ 
can be written
as
\begin{equation}
P_f^{(i)}(E,l)=\frac{1}{1+\frac{\Gamma _n}{\Gamma _f}(E,l,i)}
\label{eq:P_f}
\end{equation}
where the angular momentum dependence comes in through the rotational
energy dependence of $\Gamma _f/\Gamma _n$ and the ``multiple-chance''
energy dependence is accounted for on average by assuming that with
the emission of each neutron, the excitation energy drops by $2T+B_n$.

\begin{figure}
\centerline{\includegraphics[width=7.0cm,angle=90]{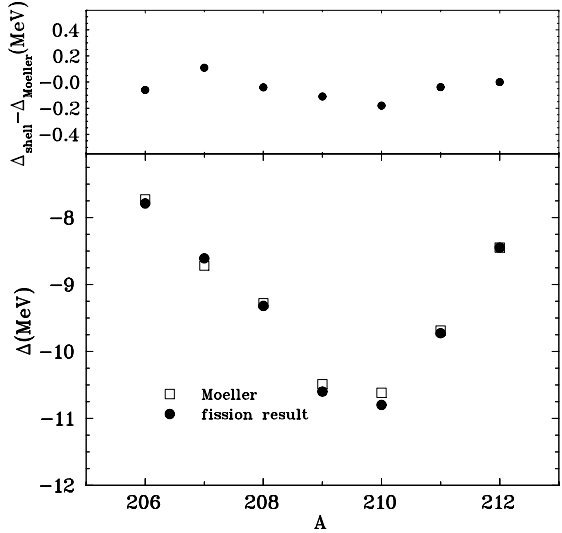}}
\caption{The shell corrections from the fission fits (solid
circles) and the shell corrections from M\"oller {\em et al.}
\protect\cite{Moller94} are plotted as a function of mass number for
Po. The residual difference between the two data sets is shown in the upper panel.}
\label{fig:shell}
\end{figure}

With Eqs.~(\ref{eq:exact}), and
(\ref{eq:gammaf_gamman})-(\ref{eq:P_f}), we are prepared to fit any
chain of neighboring isotope fission data.
However, a remark regarding the fusion cross sections is in order.
 If we use the Bass model description of the fusion cross
sections \cite{Bass} and fit the fission cross sections with the method outlined
above, we get reasonable fits to the $\alpha$-induced reactions, but
somewhat poorer fits for the $^{3}$He-induced reactions. The Bass model may not describe both the $^3$He- and $^4$He-induced fusion cross sections.
Therefore, we have chosen to leave the fusion cross section as a free parameter constrained to  the form:
\begin{equation}
\sigma_0= \frac{E_2-V}{E_{cm}}\pi R^2\tanh\left(\frac{E_{cm}-V}{E_2-V}\right)
\label{eq:myfusion}
\end{equation}
where $V$ represents the fusion barrier, $\pi R^2$ is a geometric
cross section and $E_2 = 1/2 \mu v_{\rm rel}^2$, the energy above which the fusion cross section effectively falls like 1/$E_{cm}$.  Note that in the low energy limit ($E_{cm}\approx V$),
Eq.~(\ref{eq:myfusion}) goes to
\begin{equation}
\sigma_0=\pi R^2\left(1-\frac{V}{E_{cm}}\right)
\end{equation}
and at high energies $\sigma_0$ goes to
\begin{equation}
\sigma_0=\frac{E_2-V}{E_{cm}}\pi R^2.
\end{equation}

\begin{figure}
\centerline{\includegraphics[width=7.5cm,angle=90]{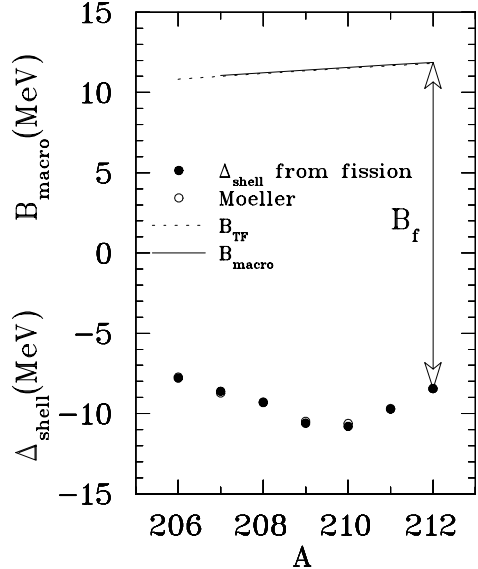}}
\caption{The shell corrections extracted from the fission fits (solid
circles) are plotted as a function of mass number. The open circles
represent the ground state shell correction estimated by M\"oller {\em
et al.} \protect\cite{Moller94}. The solid line is the macroscopic
barrier extracted from the fission fit and the dashed line is a
Thomas-Fermi estimate \protect\cite{Swiatecki99}. The difference
between the the macroscopic barrier $B_{\rm macro}$ and the shell
correction $\Delta _{\rm shell}$ is the fission barrier $B_f$.}
\label{fig:barriers}
\end{figure}

With this choice of fusion cross sections we are ready to proceed and
fit the fission cross sections. Note that this new fitting technique
requires a self-consistent global description of the data. For
example, the third chance fission for nucleus $A$ uses the same
fission barrier as the second chance fission of nucleus $A-1$, which
is the same barrier as first chance fission of nucleus $A-2$. 

The total fission cross sections calculated using
Eq.~(\ref{eq:multichance}) are shown as the solid lines in
Fig.~\ref{fig:fits}. The dashed line represents ``first-chance''
fission. The dotted lines represent second, third and higher chance
fission yields.

To fit all of the systems in Fig.~\ref{fig:fits}, eleven free parameters
were taken: three to describe the fusion cross sections ($v_{\rm rel}$ and one $R$ for each projectile type, see Eq.~(\ref{eq:myfusion})), one to describe the $A$ dependence of the macroscopic
barriers, one to describe the average value of $a_f/a_d$, and one each to describe the six
shell corrections for the 1n daughter channel of the six
fissioning systems (the shell correction for $^{212}$Po was fixed at
the M\"oller value \cite{Moller94}).

The extracted $\Delta _{\rm shell}$ values are shown by the solid
circles in Fig.~\ref{fig:shell}. They show a clear shell closure at $A=210$ $(N=126)$. Furthermore,
there is a remarkable agreement between the values from the present
fission analysis and those determined by M\"oller {\em et al.} in
fitting the ground state masses \cite{Moller94} (open squares). The
mean deviations are smaller than 200 keV (upper panel of Fig~\ref{fig:shell}). 

The agreement is remarkable, especially compared to earlier
attempts \cite{Moretto95b} 
(fitting one compound system using a
``first-chance-emission only'' formalism) where the uncertainties were $\approx\pm
1.5$ MeV. 
The errors from the present analysis suffer from a lack of exact knowledge of the fusion cross section, a value of $a_f/a_d$, and $B_{\rm macro}$. Because these three parameters are so strongly correlated, the chi square space of the fit is very flat and the resulting error matrix  is not positive-definite. Consequently, errors from the full fit cannot be assigned. However, if values for $\sigma_0$, $a_f/a_d$, and $B_{\rm macro}$ are ``frozen'' to their best values and the fits repeated with only the shell corrections free, the resulting calculated errors are less than 10 keV.  

Furthermore, the extraction of shell corrections from the fission fits 
offers an alternative way to measure shell effects which is purely local, i.e. it does not depend on the (assumed) liquid drop background.

These shell effects modify the fission probability according to Eq.~(\ref{eq:gammaf_gamman}) up to the highest excitation energies ($\approx$ 150 MeV). This observation is in accordance with the theoretical expectations of the excitation energy dependence of shell effects and at variance with recent claims of loss of shell structure at high energies \cite{Junghans98}.
 
The extracted fission barriers are shown in Fig.~\ref{fig:barriers} as
a difference between the shell correction and the macroscopic
barriers.  The macroscopic barrier from the fit is given by the solid
line and is nearly indistinguishable from the Thomas-Fermi prediction
(solid line) \cite{Swiatecki99}. With data at other fissilities, it
should be possible to explore systematic changes in the macroscopic
barriers, in particular the shape changes of the congruence energy
predicted by Myers and Swiatecki \cite{congruence}.

The ratio $a_f/a_d$ has an average value of $\approx 1.02$. With additional data at other values of
fissility, it should be possible to study the surface area 
dependence of $a_f$ \cite{Tok81}.

%

In summary, we have reported new precision fission data, and we have extracted accurate fission
barriers and ground state shell corrections with a new method of
globally fitting fission data for an isotopic chain of nuclei. An
accurate description of the saddle mass configuration may open avenues
that have been explored extensively for ground
state masses. For example, it may soon be possible to address pairing
corrections at the saddle, the surface area (or fissility) dependence
of both the saddle level density and the macroscopic barrier, and even
shell effects at the saddle in a quantitative fashion. As more data
become available, especially at the new radioactive beam facilities,
the techniques presented here may prove valuable for an
accurate description and understanding
of the fission ``saddle-mass'' surface.

\begin{acknowledgments}
This work was supported by the US Department of Energy.
\end{acknowledgments}

%
%
%



\end{document}